\documentclass[a4paper]{article}

\usepackage{a4wide}
\usepackage[UKenglish]{babel}
\usepackage{graphicx,subfigure}
	\graphicspath{{./}{figures/}}
\usepackage[colorlinks=true,linkcolor=blue,citecolor=red]{hyperref}
\usepackage{amssymb,empheq,bbold}
\usepackage[inline]{enumitem}
\usepackage{caption} 
\newcommand{\abs}[1]{\left\lvert#1\right\rvert}
\newcommand{\Cov}[2]{\operatorname{Cov}(#1,\,#2)}
\newcommand{\Prob}{\operatorname{Prob}}
\newcommand{\R}{\mathbb{R}}
\newcommand{\sgn}[1]{\operatorname{sgn}\left(#1\right)}

\begin{document}
\title{Opinion polarisation in social networks}
\author{Nadia Loy\thanks{\texttt{nadia.loy@polito.it}} \and Matteo Raviola\thanks{\texttt{s288161@studenti.polito.it}} \and Andrea Tosin\thanks{\texttt{andrea.tosin@polito.it}}}
\date{\small Department of Mathematical Sciences ``G. L. Lagrange'' \\ Politecnico di Torino, Italy}

\maketitle

\begin{abstract}
In this paper, we propose a Boltzmann-type kinetic description of opinion formation on social networks, which takes into account a general connectivity distribution of the individuals. We consider opinion exchange processes inspired by the Sznajd model and related simplifications but we do not assume that individuals interact on a regular lattice. Instead, we describe the structure of the social network statistically, assuming that the number of contacts of a given individual determines the probability that their opinion reaches and influences the opinion of another individual. From the kinetic description of the system, we study the evolution of the mean opinion, whence we find precise analytical conditions under which a \textit{polarisation switch} of the opinions, i.e. a change of sign between the initial and the asymptotic mean opinions, occurs. In particular, we show that a non-zero correlation between the initial opinions and the connectivity of the individuals is necessary to observe polarisation switch. Finally, we validate our analytical results through Monte Carlo simulations of the stochastic opinion exchange processes on the social network.

\medskip


\medskip

\end{abstract}

\section{Introduction}
Network-structured interactions permeate modern societies, a prominent example being online communication platforms. It is therefore not surprising that a large part of sociophysical studies focuses on the influence that an underlying network of connections among the individuals has on the emergence of aggregate social trends.

Early theoretical investigations were devoted to the construction of graph models for complex networks~\cite{newman2002PNAS,watts1998NATURE} and to the characterisation of the \textit{statistical} structure of social networks~\cite{albert2002RMP,barabasi1999SCIENCE,barabasi1999PHYSA}. The reason for a statistical approach is clear: typically the number of nodes and links of a social network is so large that a detailed description by means of classical graphs would be largely unfeasible. Subsequently, the interest switched to the network-structured dynamical evolution of social determinants, such as the wealth~\cite{lanchier2017JSP,lanchier2018JSP} or the opinion~\cite{acemoglu2010DGA,albi2017KRM,toscani2018PRE} of the individuals to name just the probably most common examples. Various mathematical approaches to these problems have been proposed. Without intending to review all the pertinent literature, here we simply recall some contributions relatively close to the approach that we will adopt in this paper:
\begin{enumerate}[label=\roman*)]
\item microscopic models based on tracking the time evolution of the state of every node of the network (with the identification ``node = agent'' or ``node = metapopulation'')~\cite{bertotti2016EPJST,caponigro2015DCDS,zino2020CHAOS};
\item mesoscopic models which incorporate a statistical description of the connectivity of the individuals to describe the time evolution of the distribution function of the social traits of interest~\cite{albi2017KRM,toscani2018PRE};
\item mesoscopic models, and corresponding macroscopic limits, in which the individuals are labelled by a variable discriminating their mutual interactions, which reproduces a (weighted) graph~\cite{burger2021VJM,loy2021KRM,loy2021MBE}.
\end{enumerate}

In this paper, we are interested in opinion dynamics on social networks. A powerful mathematical paradigm that has emerged in the last twenty years to address opinion formation problems is inspired by \textit{statistical mechanics} and consists in a revisitation of the methods of the \textit{collisional kinetic theory} applied to \textit{interacting multi-agent systems}~\cite{pareschi2013BOOK,toscani2006CMS}. Kinetic equations allow one to investigate rigorously the emergence of complex aggregate features, such as the opinion distribution in a human society, starting from simple heuristic descriptions of the individual interactions. However, most kinetic models of opinion formation do not consider networked interactions. They assume instead that every individual can affect the opinion of every other individual, at least within a certain opinion distance (\textit{bounded confidence}, cf.~\cite{hegselmann2002JASSS}). Instead, here we want to include the effect of the connectivity of the users of a social network on the opinion formation process, so as to investigate the impact that the distribution of contacts has on the persuasiveness and penetration of the opinions on the social network. In particular, we will opt for a statistical description of the connectivity, which can be quite naturally embedded in a kinetic framework. At the same time, in order to make the problem amenable to analytical investigations, we will simplify the opinion exchange setting, which usually assumes that the opinion is a continuous variable ranging in some bounded interval, such as e.g., $[-1,\,1]$, cf.~\cite{aletti2007SIAP,boudin2010CHAPTER,toscani2006CMS}. Taking inspiration from the Sznajd model~\cite{sznajd-weron2000IJMP,sznajd-weron2021PHYSA}, we will consider only the two opposite opinions $\pm 1$, which are conceptually analogous to the atom spins of the celebrated Ising model for the magnetisation of the matter~\cite{ising1925ZFP}.

We will investigate the appearance of what we call a \textit{polarisation switch} in the opinion distribution on the social network. By polarisation switch we mean, in particular, that the asymptotic mean opinion emerging in the long run has opposite sign with respect to the initial mean opinion, implying that most individuals switch to opposite sentiments over time. We stress that a polarisation switch is different from a \textit{phase transition}, which refers instead to a transition from disordered to ordered (or vice versa) opinion distributions over time without the possibility for the mean opinion to change sign. Concerning this, we mention that several investigations of the phase transition in the Sznajd model may be found in the literature, cf. e.g.,~\cite{calvelli2019PHYSA,muslim2020IJMP,sabatelli2003IJMP,slanina2003EPJB,sznajd-weron2011EPL,woloszny2007PHYSA}. Owing to the legacy of the Ising model, these works typically regard the network of individuals as either a complete graph or a regular lattice. Consequently, interactions follow a first-neighbour scheme and the geometrical dimension of the lattice plays a major role in determining the emergence of phase transitions. Nevertheless, social networks cannot be completely assimilated to lattices, because the latter are in general too regular and do not place enough emphasis on the possibly heterogeneous distribution of the connectivity meant as the number of contacts of the individuals. In this work, we conceive instead a quite general statistical description of the connectivity, which enters the opinion exchange process through the probability that the opinion of a certain user reaches and affects that of another user of the social network. Taking then advantage of the methods of the kinetic theory, we show that a generic connectivity distribution may allow for polarisation switch and we obtain precise analytical conditions under which the latter may occur. The conditions that we find are valid for every connectivity distribution, hence for every statistical characterisation of the social network.

In more detail, the paper is organised as follows. In Section~\ref{sect:particles_and_Boltzmann} we introduce the general kinetic approach to opinion formation on social networks, we detail two stochastic particle models at the basis of the opinion exchange schemes that we consider in the work and finally we give the corresponding kinetic descriptions in terms of Boltzmann-type equations. In Section~\ref{sect:polarisation_switch} we use the Boltzmann-type equations to investigate the emergence of polarisation switch in the two opinion exchange models previously introduced. In Section~\ref{sect:additional} we draw further conclusions about the opinion formation processes on the social network. In particular, we determine explicitly the large-time opinion distributions and we explore the importance of the statistical correlation between opinion and connectivity for the emergence of polarisation switch. In Section~\ref{sect:numerics} we simulate the original stochastic particle models via a Monte Carlo method and we compare the numerical results with the analytical predictions of the kinetic equations to validate our theoretical results. Finally, in Section~\ref{sect:conclusions} we summarise the main results of the paper and we briefly sketch possible research developments.

\section{Particle models and their Boltzmann-type descriptions}
\label{sect:particles_and_Boltzmann}
\subsection{Preliminaries}
\label{sect:preliminaries}
We consider a social network with a large number of users. We describe the state of a generic user by their opinion-connectivity pair $(w,\,c)$, where $w\in\{-1,\,1\}$ is a discrete binary variable and $c\in\R_+$ is a continuous non-negative one. The values $w=\pm 1$ denote conventionally two opposite opinions in the same spirit as the Sznajd model~\cite{sznajd-weron2000IJMP,sznajd-weron2021PHYSA}. The connectivity $c$ is assumed to be a representative measure of the \textit{followers} of a given individual, namely of the number of users who may be exposed to the opinion expressed by that individual. We consider this variable continuous in accordance with the reference literature on the connectivity distribution of social networks, see e.g.,~\cite{barabasi1999SCIENCE,barabasi1999PHYSA,clauset2009SIREV,newman2002PNAS,watts1998NATURE}.

Let $f=f(w,c,t)$ be the kinetic distribution function of the pair $(w,\,c)$ at time $t\geq 0$. Owing to the discreteness of $w$, we may represent it as
\begin{equation}
	f(w,c,t)=p(c,t)\delta(w-1)+q(c,t)\delta(w+1),
	\label{eq:f}
\end{equation}
where $\delta(w-w_0)$ denotes the Dirac delta distribution centred at $w=w_0$ and $p,\,q\geq 0$ are coefficients which depend in general on $c$ and $t$. Moreover, we assume $p(\cdot,t),\,q(\cdot,t)\in L^1(\R_+)$ for all $t\geq 0$. Considering a constant-in-time number of users and connections of the social network, we may impose the normalisation condition
\begin{equation}
	\int_{\R_+}\int_{\{-1,\,1\}}f(w,c,t)\,dw\,dc=1 \qquad \forall\,t\geq 0
	\label{eq:f.normalisation}
\end{equation}
and consequently think of $f$ as the probability density of the pair $(w,\,c)$.

The opinion density is then given by the marginal
$$ h(w,t):=\int_{\R_+}f(w,c,t)\,dc=\hat{p}(t)\delta(w-1)+\hat{q}(t)\delta(w+1), $$
where we have set
$$ \hat{p}(t):=\int_{\R_+}p(c,t)\,dc, \qquad \hat{q}(t):=\int_{\R_+}q(c,t)\,dc. $$
Notice that $\hat{p}(t)$, $\hat{q}(t)$ are the probabilities that an individual expresses the opinion $w=1$ or $w=-1$, respectively, at time $t$. Consistently with~\eqref{eq:f.normalisation}, it results
\begin{equation}
	\hat{p}(t)+\hat{q}(t)=1, \qquad \forall\,t\geq 0.
	\label{eq:phat+qhat}
\end{equation}

The connectivity density is instead given by the marginal
$$ g(c,t):=\int_{\{-1,\,1\}}f(w,c,t)\,dw=p(c,t)+q(c,t). $$
We assume that the number of followers of a generic individual possibly varies in time more slowly than their opinion, so that the marginal distribution $g$ may be well considered constant in time: $g(c,t)=g(c)$ for all $t\geq 0$. Consequently, the sum $p(c,t)+q(c,t)$ is constant in $t$ although the single terms $p(c,t)$, $q(c,t)$ may be not.

\subsection{Particle models}
We now describe two representative particle models of opinion exchange, which are at the basis of the kinetic equations satisfied by the distribution function $f$ which we will subsequently analyse.

Let $W_t\in\{-1,\,1\}$ and $C\in\R_+$ be two random variables, whose joint distribution at time $t$ is $f(w,c,t)$. They represent the opinion and connectivity, respectively, of a generic user of the social network. Consistently with the discussion set forth at the end of Section~\ref{sect:preliminaries}, $C$ is constant in time. Conversely, $\{W_t,\,t\in [0,\,+\infty)\}$ is the stochastic process of opinion formation of the given user.

Taking inspiration from the opinion exchange models presented in~\cite{slanina2003EPJB}, which are in turn revisitations of the Sznajd model~\cite{sznajd-weron2000IJMP}, see also~\cite{stauffer2000IJMP}, we consider the following interaction schemes:
\begin{enumerate}[label=\roman*)]
\item The \textbf{two-against-one} model, which assumes \textit{ternary} interactions in which the third individual takes the opinion of the first two ones if the latter have the same opinion; otherwise, no interaction takes place. In formulas:
\begin{equation}
\begin{cases}
	W_{t+\Delta{t}}=W_t \\
	W^\ast_{t+\Delta{t}}=W^\ast_t \\
	W^{\ast\ast}_{t+\Delta{t}}=(1-\Theta)W^{\ast\ast}_t+\Theta W_t,
\end{cases}
\label{eq:2-vs-1.particle}
\end{equation}
where $W^\ast_t,\,W^{\ast\ast}_t\in\{-1,\,1\}$ are the opinions of two further users of the social network and $\Delta{t}>0$ is the amplitude of the time interval in which the interaction may happen. Moreover, $\Theta\in\{0,\,1\}$ is a Bernoulli random variable discriminating whether the interaction takes place ($\Theta=1$) or not ($\Theta=0$). We assume:
\begin{equation}
	\Prob(\Theta=1):=\mu\chi(W_t=W^\ast_t)CC^\ast\Delta{t},
	\label{eq:2-vs-1.Theta}
\end{equation}
where $\mu>0$ is a proportionality constant and $\chi(\cdot)$ denotes the characteristic function of the event indicated in parenthesis. Hence, consistently with the discussion above, an interaction may take place only if $W_t=W^\ast_t$. In such a case, the probability that the third individual is reached by the common opinion of the first two individuals and gets convinced by them is proportional to the connectivities of the first two individuals and to the duration $\Delta{t}$ of the interaction interval. Notice that, for consistency, in~\eqref{eq:2-vs-1.Theta} we need to assume
$$ \Delta{t}\leq\frac{1}{\mu\max{(CC^\ast)}}, $$
the maximum being taken over all individuals of the system.
\item An \textbf{Ochrombel-type simplification} of~\eqref{eq:2-vs-1.particle}, cf.~\cite{ochrombel2001IJMP}, which assumes that a cluster of identical opinions (in our case, two identical opinions) is \textit{not} necessary to convince a further individual. Instead, any individual is in principle able to convince another individual in a \textit{binary} interaction, so that the particle model becomes:
\begin{equation}
\begin{cases}
	W_{t+\Delta{t}}=W_t \\
	W^\ast_{t+\Delta{t}}=(1-\Theta)W^\ast_t+\Theta W_t,
\end{cases}
\label{eq:Ochrombel.particle}
\end{equation}
where this time we set
\begin{equation}
	\Prob(\Theta=1):=\mu C\Delta{t}
	\label{eq:Ochrombel.Theta}
\end{equation}
to reproduce the idea that the probability that the second individual is reached and convinced by the opinion of the first individual is proportional to the connectivity of the latter and to the duration of the interaction. For consistency we assume
$$ \Delta{t}\leq\frac{1}{\mu\max{C}}, $$
the maximum being again taken over all individuals of the system.
\end{enumerate}

\subsection{Boltzmann-type descriptions}
Following standard procedures, see e.g.,~\cite{pareschi2013BOOK} or~\cite[Appendix A]{fraia2020RUMI}, the discrete-in-time stochastic particle models~\eqref{eq:2-vs-1.particle},~\eqref{eq:Ochrombel.particle} may be given a continuous-in-time statistical description in the limit $\Delta{t}\to 0^+$ in terms of Boltzmann-type ``collisional'' equations for the distribution function $f$.

The two-against-one model~\eqref{eq:2-vs-1.particle} involves interactions among three individuals at a time, thus it requires a multiple-interaction kinetic equation, cf.~\cite{bobylev2009CMP,toscani2020NHM}. In this context ``multiple'' means ``more than pairwise'', interactions in pairs being the common standard in kinetic theory. In weak form, using an arbitrary \textit{observable quantity} (test function) $\phi=\phi(w,c):\{-1,\,1\}\times\R_+\to\R$, the multiple-interaction kinetic equation describing the evolution of $f$ ruled by the particle model~\eqref{eq:2-vs-1.particle} reads
\begin{multline}
	\frac{d}{dt}\int_{\R_+}\int_{\{-1,\,1\}}\phi(w,c)f(w,c,t)\,dw\,dc \\
	=\frac{1}{3}\int_{\R_+^3}\int_{\{-1,\,1\}^3}B(w,c,w_\ast,c_\ast)\bigl(\phi(w,c_{\ast\ast})-\phi(w_{\ast\ast},c_{\ast\ast})\bigr) \\
		\times f(w,c,t)f(w_\ast,c_\ast,t)f(w_{\ast\ast},c_{\ast\ast},t)\,dw\,dw_\ast\,dw_{\ast\ast}\,dc\,dc_\ast\,dc_{\ast\ast},
	\label{eq:2-vs-1.Boltz}
\end{multline}
where the \textit{collision kernel} $B$ is
$$ B(w,c,w_\ast,c_\ast):=\mu\chi(w=w_\ast)cc_\ast. $$
This is the interaction frequency induced by the choice~\eqref{eq:2-vs-1.Theta} of the law of $\Theta$. Notice that such a collision kernel confers on the kinetic equation~\eqref{eq:2-vs-1.Boltz} a \textit{non-Maxwellian} character with \textit{cut-off}, because $B$ is non-constant and the term $\chi(w=w_\ast)$ excludes the interactions with $w\neq w_\ast$.

The Ochrombel-type simplification~\eqref{eq:Ochrombel.particle} features instead binary interactions, hence the corresponding equation for $f$ resembles more closely the classical kinetic equations of statistical mechanics:
\begin{multline}
	\frac{d}{dt}\int_{\R_+}\int_{\{-1,\,1\}}\phi(w,c)f(w,c,t)\,dw\,dc \\
	=\frac{1}{2}\int_{\R_+^2}\int_{\{-1,\,1\}^2}B(c)\bigl(\phi(w,c_\ast)-\phi(w_\ast,c_\ast)\bigr)f(w,c,t)f(w_\ast,c_\ast,t)\,dw\,dw_\ast\,dc\,dc_\ast.
	\label{eq:Ochrombel.Boltz}
\end{multline}
In this case, the collision kernel is
$$ B(c):=\mu c, $$
consistently with the interaction frequency induced by the choice~\eqref{eq:Ochrombel.Theta} of the law of $\Theta$. This kernel is again non-Maxwellian, because it is non-constant, but without cut-off.

\section{Conditions for polarisation switch}
\label{sect:polarisation_switch}
We say that the opinion formation process on the social network exhibits a \textit{polarisation switch} if the initial mean opinion of the users and their asymptotic mean opinion, i.e. the mean opinion emerging in the long run in consequence of the interactions, have opposite sign. The mean opinion is defined as
\begin{equation}
	m_W(t):=\int_{\R_+}\int_{\{-1,\,1\}}wf(w,c,t)\,dw\,dc=\int_{\{-1,\,1\}}wh(w,t)\,dw=\hat{p}(t)-\hat{q}(t)
	\label{eq:mW}
\end{equation}
and in this context plays the role of the \textit{magnetisation} of the Ising model~\cite{ising1925ZFP}, cf. also~\cite{slanina2003EPJB}. Denoting by $m_W^0$ and $m_W^\infty$ the initial and asymptotic mean opinions, respectively, the condition for polarisation switch may be expressed as
$$ m_W^0m_W^\infty<0. $$
Notice that, according to this definition, if $m_W^0=0$ we cannot speak of polarisation switch regardless of $m_W^\infty$. Assuming therefore $m_W^0\neq 0$, an equivalent condition for polarisation switch that we will use in the sequel is
\begin{equation}
	\frac{m_W^\infty}{m_W^0}<0.
	\label{eq:phase_trans}
\end{equation}

In this section, we will establish precise conditions for polarisation switch to happen in terms of statistical features of the connectivity of the social network jointly with some aggregate characteristics of the initial distribution of the opinions.

\subsection{The two-against-one model}
\label{sect:2-vs-1}
To study the time evolution of $m_W$ towards $m_W^\infty$ in model~\eqref{eq:2-vs-1.particle} we choose $\phi(w,c)=w$ in~\eqref{eq:2-vs-1.Boltz} and we take advantage of the representation~\eqref{eq:f} of the distribution function $f$. After some computations we get:
$$ \dot{m}_W(t)=\frac{2}{3}\mu\left[{\left(\int_{\R_+}cp(c,t)\,dc\right)}^2\hat{q}(t)
	-{\left(\int_{\R_+}cq(c,t)\,dc\right)}^2\hat{p}(t)\right]. $$
Next, from~\eqref{eq:phat+qhat} and~\eqref{eq:mW} we observe that
$$ \hat{p}(t)=\frac{1+m_W(t)}{2}, \qquad \hat{q}(t)=\frac{1-m_W(t)}{2}. $$
Moreover, by introducing the \textit{product moment} $m_{WC}$ defined as
\begin{equation}
	m_{WC}(t):=\int_{\R_+}\int_{\{-1,\,1\}}wcf(w,c,t)\,dw\,dc=\int_{\R_+}cp(c,t)\,dc-\int_{\R_+}cq(c,t)\,dc
	\label{eq:mWC}
\end{equation}
and the mean connectivity
$$ m_C:=\int_{\R_+}cg(c)\,dc=\int_{\R_+}cp(c,t)\,dc+\int_{\R_+}cq(c,t)\,dc $$
we deduce
$$ \int_{\R_+}cp(c,t)\,dc=\frac{m_C+m_{WC}(t)}{2}, \qquad \int_{\R_+}cq(c,t)\,dc=\frac{m_C-m_{WC}(t)}{2}, $$
whence, after some algebraic manipulations, we rewrite the equation for $m_W$ in the form
\begin{equation}
	\dot{m}_W=\frac{\mu}{6}\left[2m_Cm_{WC}-\left(m_C^2+m_{WC}^2\right)m_W\right].
	\label{eq:2-vs-1.mW}
\end{equation}
Notice that $m_C$ is constant in time, because so is the entire marginal distribution $g$ of the connectivity, and may therefore be considered as known once the characteristics of the network are fixed. In particular, throughout the paper we will assume $m_C>0$, for $m_C=0$ would imply $g(c)=\delta(c)$, i.e. no social connections at all. Conversely, $m_{WC}$ is in general not constant, hence~\eqref{eq:2-vs-1.mW} is not sufficient by itself to extract information on the large time trend of $m_W$.

From~\eqref{eq:2-vs-1.Boltz} we can study the evolution of $m_{WC}$ by choosing $\phi(w,c)=wc$. This gives, after some computations,
\begin{equation}
	\dot{m}_{WC}=\frac{\mu}{6}\left(m_C^2-m_{WC}^2\right)m_{WC},
	\label{eq:2-vs-1.mWC}
\end{equation}
which is a self-consistent equation for $m_{WC}$. Solving by separation of variables we find
$$ m_{WC}(t)=\frac{m_{WC}^0m_Ce^{\frac{\mu}{6}m_C^2t}}{\sqrt{m_C^2+{(m_{WC}^0)}^2\left(e^{\frac{\mu}{3}m_C^2t}-1\right)}}, $$
where $m_{WC}^0$ denotes the initial value of the product moment. From this representation formula we obtain in particular that:
\begin{enumerate}[label=\roman*)]
\item $m_{WC}(t)\to m_{WC}^\infty:=\sgn{m_{WC}^0}m_C$ for $t\to+\infty$, thus $m_{WC}$ reaches asymptotically the values $\pm m_C$ depending on whether it is initially positive or negative. If instead $m_{WC}^0=0$ then $m_{WC}$ remains zero at all times;
\item the convergence of $m_{WC}(t)$ to $m_{WC}^\infty=\sgn{m_{WC}^0}m_C$ is exponentially fast in time, indeed:
\begin{equation}
	\abs{m_{WC}(t)-m_{WC}^\infty}\leq m_C\abs{\frac{e^{\frac{\mu}{6}m_C^2t}}{\sqrt{e^{\frac{\mu}{3}m_C^2t}-1}}-1}\sim\frac{m_C}{2}e^{-\frac{\mu}{3}m_C^2t}
		\quad \text{for } t\to+\infty.
	\label{eq:mWC.limit}
\end{equation}
\end{enumerate}

These two facts allow us to infer from~\eqref{eq:2-vs-1.mW} the asymptotic trend of $m_W$. By rewriting~\eqref{eq:2-vs-1.mW} as
$$ \dot{m}_W=\frac{\mu}{6}\left(m_C^2+m_{WC}^2\right)\left(\frac{2m_Cm_{WC}}{m_C^2+m_{WC}^2}-m_W\right), $$
we observe that it is an ordinary differential equation of the form
$$ \dot{x}=a(t)(b(t)-x). $$
It is known that if $a(t)$ is such that $\underline{a}\leq a(t)\leq\overline{a}$ for two constants $\underline{a},\,\overline{a}>0$ and if $b(t)$ converges exponentially fast to a limit value $b^\infty$ when $t\to+\infty$, i.e. $\abs{b(t)-b^\infty}\lesssim e^{-\tilde{b}t}$ for $t$ sufficiently large and for a certain $\tilde{b}>\overline{a}-\underline{a}$, then also $x$ converges to $b^\infty$ as $t\to+\infty$. In our case, we have $x(t)=m_W(t)$ and
$$ a(t)=\frac{\mu}{6}\left(m_C^2+m_{WC}^2(t)\right), \qquad b(t)=\frac{2m_Cm_{WC}(t)}{m_C^2+m_{WC}^2(t)}. $$
From~\eqref{eq:mWC} it results in general $\abs{m_{WC}(t)}\leq m_C$, thus $\underline{a}:=\frac{\mu}{6}m_C^2\leq a(t)\leq\frac{\mu}{3}m_C^2=:\overline{a}$. Furthermore, $b(t)\to\sgn{m_{WC}^0}=:b^\infty$ when $t\to+\infty$ and
\begin{align*}
	\abs{b(t)-b^\infty} &= \abs{\frac{2m_Cm_{WC}}{m_C^2+m_{WC}^2}-\frac{2m_Cm_{WC}^\infty}{m_C^2+{(m_{WC}^\infty)}^2}} \\
	&= 2m_C\frac{\abs{m_{WC}\left(m_C^2+{(m_{WC}^\infty)}^2\right)-m_{WC}^\infty\left(m_C^2+m_{WC}^2\right)}}{\left(m_C^2+m_{WC}^2\right)\left(m_C^2+{(m_{WC}^\infty)}^2\right)} \\
	&\leq \frac{2}{m_C}\left(1+\frac{1}{m_C}\right)\abs{m_{WC}-m_{WC}^\infty},
\end{align*}
which, owing to~\eqref{eq:mWC.limit}, converges to zero exponentially fast when $t\to+\infty$ with $\tilde{b}:=\frac{\mu}{3}m_C^2$. Since
$$ \tilde{b}=\frac{\mu}{3}m_C^2>\frac{\mu}{6}m_C^2=\overline{a}-\underline{a}, $$
we conclude
$$ m_W(t)\to m_W^\infty:=\sgn{m_{WC}^0} \quad \text{for } t\to+\infty. $$

This result may be rewritten in a more informative form considering that
$$ m_{WC}(t)=m_Cm_W(t)+\Cov{W_t}{C}, $$
$\Cov{\cdot}{\cdot}$ being the covariance, and that $\sgn{m_{WC}^0}=\sgn{\frac{m_{WC}^0}{m_C}}$. In particular,
$$ \frac{m_{WC}^0}{m_C}=m_W^0+\frac{\Cov{W_0}{C}}{m_C}=m_W^0+\frac{\rho_{WC}^0\sigma_W^0\sigma_C}{m_C}, $$
where $\rho_{WC}^0$ is the correlation coefficient between the random variables $W_0$ and $C$ and $\sigma_W^0$, $\sigma_C$ are their respective standard deviations. Observing furthermore that
$$ \sigma_W^0=\sqrt{\int_{\{-1,\,1\}}w^2h(w,0)\,dw-{(m_W^0)}^2}=\sqrt{\hat{p}(0)+\hat{q}(0)-{(m_W^0)}^2}=\sqrt{1-{(m_W^0)}^2}, $$
where in the last passage we have recalled~\eqref{eq:phat+qhat}, we conclude
$$ m_W^\infty=\sgn{\frac{m_{WC}^0}{m_C}}=\sgn{m_W^0+\frac{\sigma_C}{m_C}\rho_{WC}^0\sqrt{1-{(m_W^0)}^2}}, $$
hence, owing to~\eqref{eq:phase_trans}, there is polarisation switch if
$$ 0>\frac{m_W^\infty}{m_W^0}=\frac{1}{\abs{m_W^0}}\sgn{1+\frac{\sigma_C}{m_C}\rho_{WC}^0\frac{\sqrt{1-{(m_W^0)}^2}}{m_W^0}}, $$
namely if $1+\frac{\sigma_C}{m_C}\rho_{WC}^0\frac{\sqrt{1-{(m_W^0)}^2}}{m_W^0}<0$ and finally
\begin{equation}
	\begin{cases}
		\rho_{WC}^0<-\dfrac{m_C}{\sigma_C}\cdot\dfrac{m_W^0}{\sqrt{1-{(m_W^0)}^2}} & \text{if } m_W^0>0 \\[6mm]
		\rho_{WC}^0>-\dfrac{m_C}{\sigma_C}\cdot\dfrac{m_W^0}{\sqrt{1-{(m_W^0)}^2}} & \text{if } m_W^0<0.
	\end{cases}
	\label{eq:2-vs-1.phase_trans}
\end{equation}

The qualitative interpretation of this result is clear: in order for a polarisation switch to emerge in the social network, the initial correlation between opinions and connectivity must have a sign opposite to that of the initial mean opinion. Notice indeed that in~\eqref{eq:2-vs-1.phase_trans} it results $\rho_{WC}^0<0$ when $m_W^0>0$ and vice versa. This implies, in particular, that the most connected individuals, viz. the \textit{influencers} in the jargon of social networks, should express initially an opinion opposite to the mean one.

The result~\eqref{eq:2-vs-1.phase_trans} establishes quantitatively the necessary minimum threshold of positive or negative initial correlation, showing that it depends on both aggregate characteristics of the network (the mean and standard deviation of the connectivity) and the initial mean opinion itself. Notice, in particular, that the more $m_W^0$ is biased towards $\pm 1$ the higher (in absolute value) such a threshold is, consistently with the intuitive idea that it is more difficult to produce a polarisation switch in the opinions of a strongly polarised society.

\subsection{The Ochrombel-type simplification}
Repeating the same arguments as in Section~\ref{sect:2-vs-1} for the particle model~\eqref{eq:Ochrombel.particle}, we obtain from the corresponding kinetic equation~\eqref{eq:Ochrombel.Boltz} with $\phi(w,c)=w$ the following evolution equation for the mean opinion:
$$ \dot{m}_W=\frac{\mu}{2}\left(m_{WC}-m_Cm_W\right), $$
which again requires some additional information on the evolution of the product moment $m_{WC}$. This may be obtained by plugging $\phi(w,c)=wc$ into~\eqref{eq:Ochrombel.Boltz}, which, after some computations, yields
$$ \dot{m}_{WC}=0. $$
Hence in the Ochrombel-type simplification $m_{WC}$ is constant in time, i.e. $m_{WC}(t)=m_{WC}^0$ for all $t>0$, which reduces the equation for $m_W$ to
$$ \dot{m}_W=\frac{\mu}{2}\left(m_{WC}^0-m_Cm_W\right). $$
The solution issuing from a given initial mean opinion $m_W^0$ is easily found as
\begin{equation}
	m_W(t)=e^{-\frac{\mu}{2}m_Ct}m_W^0+\frac{m_{WC}^0}{m_C}\left(1-e^{-\frac{\mu}{2}m_Ct}\right),
	\label{eq:Ochrombel.mW}
\end{equation}
therefore
$$ m_W(t)\to m_W^\infty:=\frac{m_{WC}^0}{m_C} \quad \text{for } t\to +\infty. $$

We observe that, apart from the sign function, this is the same quantity characterising the asymptotic mean opinion of the two-against-one model. Taking advantage of the computations performed in Section~\ref{sect:2-vs-1}, we may therefore write condition~\eqref{eq:phase_trans} for polarisation switch as
$$ 0>\frac{m_W^\infty}{m_W^0}=1+\frac{\sigma_C}{m_C}\rho_{WC}^0\frac{\sqrt{1-{(m_W^0)}^2}}{m_W^0}, $$
which yields again~\eqref{eq:2-vs-1.phase_trans}. In conclusion, as far as the description of the emergence of polarisation switch is concerned the Ochrombel-type simplification~\eqref{eq:Ochrombel.particle} retains all the essential features of the more elaborated two-against-one model~\eqref{eq:2-vs-1.particle}.

\section{Additional considerations}
\label{sect:additional}
\subsection{Asymptotic opinion distributions}
The two-against-one model produces $m_W^\infty=\sgn{m_{WC}^0}$, hence, independently of the polarisation switch, only three values are possible for the asymptotic mean opinion: $m_W^\infty=-1,\,0,\,1$. In particular, we observe that $m_W^\infty=0$ arises only if $m_{WC}^0=0$ and that the latter is an unstable equilibrium of the product moment $m_{WC}$, cf.~\eqref{eq:2-vs-1.mWC}. As a matter of fact, the relevant physically observable cases are therefore $m_W^\infty=\pm 1$, which identify a \textit{consensus} in the social network with asymptotic opinion distribution
$$ h^\infty(w)=\delta(w\pm 1). $$

Conversely, the Ochrombel-type simplification produces $m_W^\infty=\frac{m_{WC}^0}{m_C}\in [-1,\,1]$, which need not imply an asymptotic consensus because $m_W^\infty$ may be in principle any value in the interval $[-1,\,1]$. A simple computation shows that the asymptotic variance $\sigma_W^{2,\infty}$ of the opinion is
$$ \sigma_W^{2,\infty}=1-\frac{(m_{WC}^0)^2}{m_C^2} $$
and that the asymptotic opinion distribution is in this case
\begin{equation}
	h^\infty(w)=\frac{1}{2}\left(1+\frac{m_{WC}^0}{m_C}\right)\delta(w-1)+\frac{1}{2}\left(1-\frac{m_{WC}^0}{m_C}\right)\delta(w+1).
	\label{eq:Ochrombel.hinf}
\end{equation}
All in all, the Ochrombel-type simplification produces a less sharp, thus probably more realistic, big picture of the possible asymptotic scenarios on the social network while retaining all the essential features characterising the polarisation switch.

\subsection{Statistical independence}
\label{sect:stat.indep}
An intriguing simplification of the dynamics studied in Section~\ref{sect:polarisation_switch} is obtained by assuming \textit{statistical independence} of the variables $W_t$, $C$, meaning that
\begin{equation}
	f(w,c,t)=g(c)h(w,t) \qquad \forall\,t\geq 0.
	\label{eq:stat_indip.anstaz}
\end{equation}

Plugging this ansatz into~\eqref{eq:2-vs-1.Boltz} and choosing the observable quantity of the form $\phi(w,c)=\varphi(w)\psi(c)$ for arbitrary functions $\varphi:\{-1,\,1\}\to\R$ and $\psi:\R_+\to\R$, we obtain that the Boltzmann-type equation of the two-against-one model reduces to
$$ \hat{p}'(t)\varphi(1)+\hat{q}'(t)\varphi(-1)=\frac{\mu}{3}m_C^2\hat{p}(t)\hat{q}(t)\bigl(\hat{p}(t)-\hat{q}(t)\bigr)\bigr(\varphi(1)-\varphi(-1)\bigl). $$
Recalling~\eqref{eq:phat+qhat} and invoking the arbitrariness of $\varphi$ yields
$$ \hat{p}'=\frac{\mu}{3}m_C^2\hat{p}(1-\hat{p})(2\hat{p}-1), $$
which admits the asymptotic states $\hat{p}^\infty=0,\,1$ (both stable) and $\hat{p}^\infty=\frac{1}{2}$ (unstable). To them there correspond the stable asymptotic opinion distributions
$$ h^\infty(w)=\delta(w\pm 1) $$
and the unstable one
$$ h^\infty(w)=\frac{1}{2}\delta(w-1)+\frac{1}{2}\delta(w+1), $$
which confirm that the two-against-one model tends to give rise to consensus. Moreover, the evolution of the mean opinion $m_W(t)=\hat{p}(t)-\hat{q}(t)=2\hat{p}(t)-1$ is ruled by
$$ \dot{m}_W=\frac{\mu}{6}m_C^2\left(1-m_W^2\right)m_W, $$
which, solving by separation of variables, gives
$$ m_W(t)=\frac{m_W^0}{\sqrt{{(m_W^0)}^2+\left(1-{(m_W^0)}^2\right)e^{-\frac{\mu}{3}m_C^2t}}}. $$
In particular, it results $m_W(t)\to m_W^\infty:=\sgn{m_W^0}$ as $t\to +\infty$, which shows that polarisation switch is instead never observed in this case because the asymptotic and initial mean opinions have always the same sign.

Plugging instead the ansatz~\eqref{eq:stat_indip.anstaz} into~\eqref{eq:Ochrombel.Boltz} and letting again $\phi(w,c)=\varphi(w)\psi(c)$ we find that the Boltzmann-type equation of the Ochrombel-type simplification reads
$$ \hat{p}'(t)\varphi(1)+\hat{q}'(t)\varphi(-1)=0, $$
i.e., for the arbitrariness of $\varphi$, $\hat{p}'(t)=\hat{q}'(t)=0$. Therefore, the kinetic distribution function $f$ and in particular the opinion distribution $h$ are constant in time. As a consequence, we neither observe polarisation switch nor, more in general, any modification of the statistical distribution of the opinions with respect to the initial condition.

These results are consistent with those found in~\cite{fraia2020RUMI}, where the two-against-one model and its Ochrombel-type simplification are addressed without social network, in particular by assuming that any individual may be equally reached and convinced by the opinion of any other individual regardless of the connectivity. In essence, these results show that, in the long run, the statistical independence between opinion and connectivity is equivalent to the absence of the social network. Moreover, they further stress the importance that the connectivity correlates with the expressed opinions to observe interesting aggregate dynamics including polarisation switch.

\section{Comparison with numerical simulations}
\label{sect:numerics}
In this section, we solve numerically the stochastic particle models~\eqref{eq:2-vs-1.particle}-\eqref{eq:2-vs-1.Theta} and~\eqref{eq:Ochrombel.particle}-\eqref{eq:Ochrombel.Theta} by means of a classical Monte Carlo algorithm, cf. e.g.,~\cite{pareschi2013BOOK}, and compare the outcomes of the simulations with the theoretical predictions obtained from the kinetic equations. Our numerical tests do not only provide further insights into the application considered in this paper but constitute also a genuine microscopic validation of the aggregate analytical results.

As initial condition, we consider in both cases a joint opinion-connectivity probability distribution of the form
\begin{equation}
	f^0(w,c)=\lambda K_{\alpha_p,\beta_p}(c)\delta(w-1)+(1-\lambda)K_{\alpha_q,\beta_q}(c)\delta(w+1),
	\label{eq:f0}
\end{equation}
where:
\begin{enumerate*}[label=\roman*)]
\item $\lambda\in [0,\,1]$ is the percentage of individuals expressing initially the opinion $w=1$;
\item $1-\lambda\in [0,\,1]$ is the percentage of individuals expressing initially the opinion $w=-1$;
\item $K_{\alpha,\beta}(c)$ is a two-parameter probability density function modelling the connectivity distribution of the former individuals for $\alpha=\alpha_p$, $\beta=\beta_p$ and of the latter individuals for $\alpha=\alpha_q$, $\beta=\beta_q$.
\end{enumerate*}

Following the literature, according to which many large networks feature a power-law distribution of the connectivity, cf. e.g.,~\cite{barabasi1999SCIENCE}, we choose $K_{\alpha,\beta}$ to be an inverse-gamma distribution:
$$ K_{\alpha,\beta}(c)=\frac{\beta^\alpha}{\Gamma(\alpha)}\cdot\frac{e^{-\frac{\beta}{c}}}{c^{1+\alpha}} $$
with $\alpha,\,\beta>0$ the shape and scale parameters, respectively. Notice that $K_{\alpha,\beta}(c)\sim\frac{\beta^\alpha}{\Gamma(\alpha)}c^{-(1+\alpha)}$ for $c\to +\infty$, thus for $c$ large the decay to zero obeys a power law with exponent $1+\alpha$. We may argue that the parameter $\alpha$ plays here the role of a \textit{Pareto index}~\cite{gualandi2018ECONOMICS} measuring the heaviness of the tail of $K_{\alpha,\beta}$: the lower $\alpha$ the heavier the tail, meaning that users with a high number of contacts are more frequent in the social network. In our application, these users represent the influencers.

From~\eqref{eq:f0} we deduce that the initial opinion distribution is
$$ h^0(w)=\lambda\delta(w-1)+(1-\lambda)\delta(w+1), $$
i.e. $\Prob(W_0=1)=\lambda$, $\Prob(W_0=-1)=1-\lambda$ consistently with the meaning of $\lambda$ introduced above. We also deduce that
$$ p^0(c)=\lambda K_{\alpha_p,\beta_p}(c), \qquad q^0(c)=(1-\lambda)K_{\alpha_q,\beta_q}(c), $$
hence that
$$ g(c)=\lambda K_{\alpha_p,\beta_p}(c)+(1-\lambda)K_{\alpha_q,\beta_q}(c). $$
Notice that if $\alpha_p\neq\alpha_q$ or $\beta_p\neq\beta_q$ the kinetic distribution function $f^0$ is not the product of $g$ and $h^0$, thus the opinion and the connectivity are not statistically independent.

\begin{table}[!t]
\centering
\caption{Parameters used in the numerical tests of Section~\ref{sect:numerics}}
\label{tab:parameters}
\begin{tabular}{l|ccccc}
Parameter & $\lambda$ & $\alpha_p$ & $\beta_p$ & $\beta_q$ & $\mu$ \\
\hline
\hline
Value & $0.7$ & $5$ & $300$ & $300$ & $1$ \\
\hline
\end{tabular}
\end{table}

In our numerical tests we fix the parameters listed in Table~\ref{tab:parameters}. They imply that $70\%$ of the users of the social network expresses initially the opinion $w=1$, which becomes the dominant one, with $m_W^0=0.4>0$ and $\sigma_W^0\approx 0.9$. The mean connectivity of the individuals expressing initially the dominant opinion is
$$ m_{C,p}=\frac{1}{\lambda}\int_{\R_+}cp^0(c)\,dc=\int_{\R_+}cK_{5,300}(c)\,dc=75, $$
while that of the individuals expressing initially the opinion $w=-1$ is
$$ m_{C,q}=\frac{1}{1-\lambda}\int_{\R_+}cq^0(c)\,dc=\int_{\R_+}cK_{\alpha_q,300}(c)\,dc=\frac{300}{\alpha_q-1} \qquad (\text{for }\alpha_q>1) $$
from the known formulas of the statistical moments of an inverse-gamma distribution. Furthermore, the global mean connectivity on the social network is
$$ m_C=\int_{\R_+}cg(c)\,dc=\lambda m_{C,p}+(1-\lambda)m_{C,q}=\frac{15}{2}\cdot\frac{7\alpha_q+5}{\alpha_q-1}  \qquad (\text{for }\alpha_q>1) $$
with standard deviation
$$ \sigma_C=\sqrt{\int_{\R_+}c^2g(c)\,dc-m_C^2}=\sqrt{\lambda\sigma_{C,p}^2+(1-\lambda)\sigma_{C,q}^2+\lambda(1-\lambda){(m_{C,p}-m_{C,q})}^2} $$
and
$$ \sigma_{C,p}^2=1875, \qquad \sigma_{C,q}^2=\frac{9\cdot 10^4}{{(\alpha_q-1)}^2(\alpha_q-2)} \quad (\text{for }\alpha_q>2) $$
again from the formulas of the moments of an inverse-gamma distribution.

\begin{figure}[!t]
\centering
\subfigure[]{\includegraphics[width=0.45\textwidth]{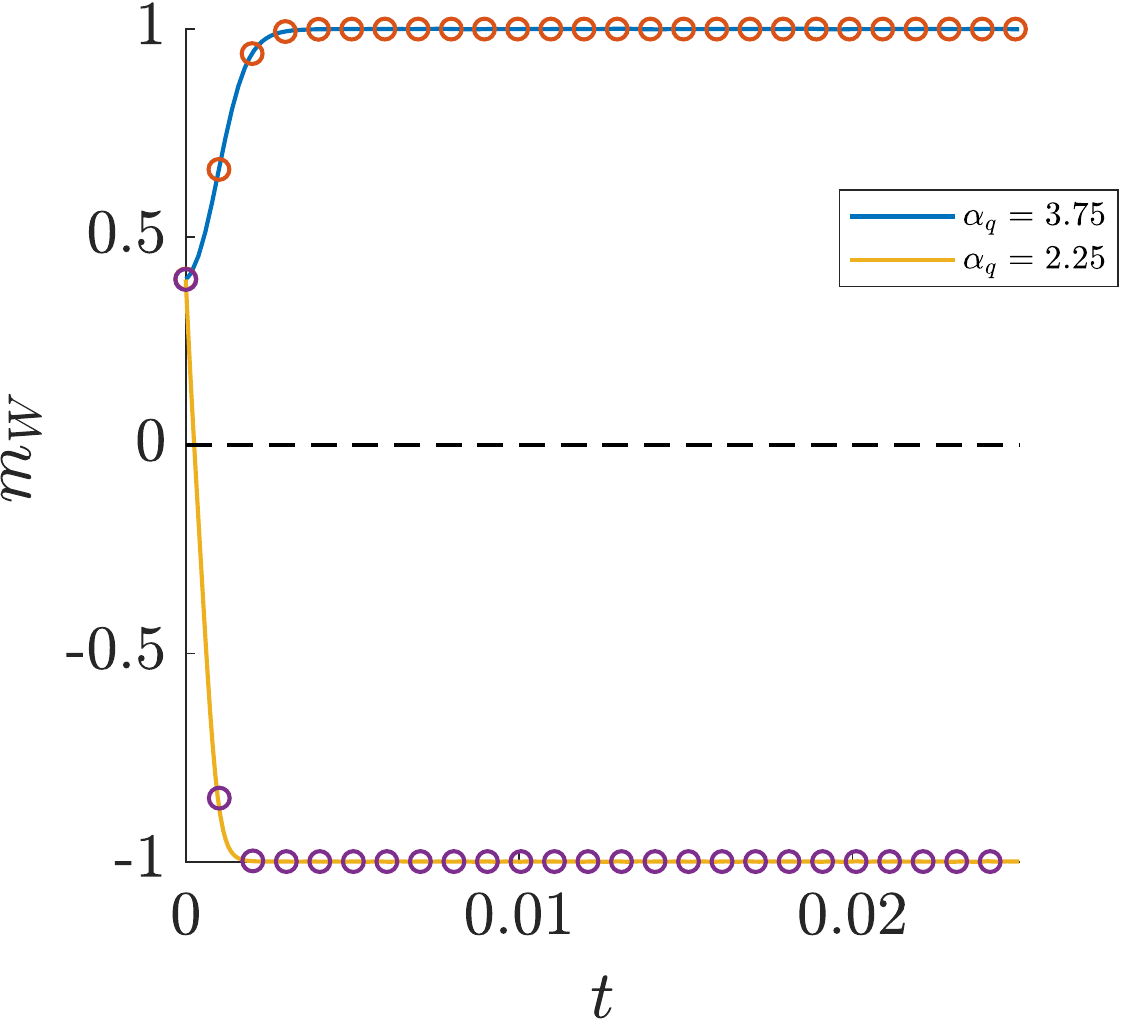}} \quad
\subfigure[]{\includegraphics[width=0.45\textwidth]{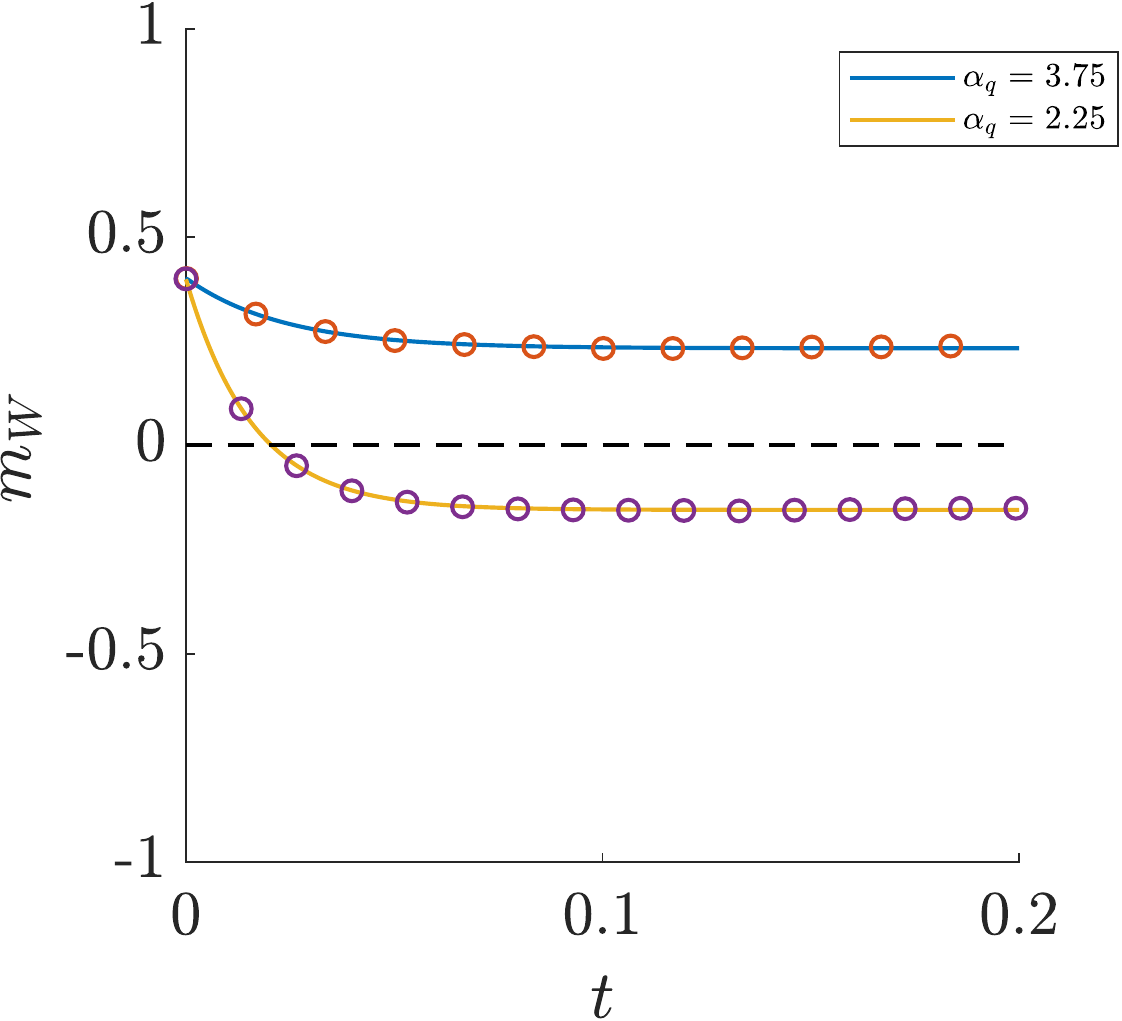}}
\caption{Time trend of the mean opinion in: (a) the two-against-one model; (b) the Ochrombel simplification for the values of the parameters in Table~\ref{tab:parameters}. Markers are the means of the Monte Carlo solutions of the stochastic particle models; solid lines are the solutions of the differential equations for $m_W$ obtained from the kinetic description.}
\label{fig:mean}
\end{figure}

We test two scenarios corresponding to the values $\alpha_q=3.75$ and $\alpha_q=2.25$:
\begin{enumerate}[label=\roman*)]
\item For $\alpha_q=3.75$ we obtain $m_{C,q}\approx 109.1$, $m_C\approx 85.2$ and $\sigma_C\approx 60$ with an initial correlation between opinion and connectivity of
$$ \rho_{WC}^0=\frac{\Cov{W_0}{C}}{\sigma_C\sigma_W^0}=\frac{m_{WC}^0-m_Cm_W^0}{\sigma_C\sigma_W^0}=\frac{\lambda m_{C,p}-(1-\lambda)m_{C,q}-m_Cm_W^0}{\sigma_C\sigma_W^0}\approx -0.27 $$
Since $m_W^0>0$, from the first condition in~\eqref{eq:2-vs-1.phase_trans} we discover that the emergence of polarisation switch would require $\rho_{WC}^0\approx-0.27<-\frac{m_C}{\sigma_C}\cdot\frac{m_W^0}{\sqrt{1-{(m_W^0)}^2}}\approx -0.62$, which is clearly violated. Therefore, in this case we do not observe polarisation switch either in the two-against-one model, cf. Figure~\ref{fig:mean}a, or in the Ochrombel-type simplification, cf. Figure~\ref{fig:mean}b. The reason is that the Pareto index $\alpha_q$ is not small enough to guarantee a sufficient presence of influencers among the individuals expressing initially the opinion $w=-1$ opposite to the dominant one. This is further stressed by the mean connectivity $m_{C,q}$ of the latter, which is only slightly greater than that of the individuals expressing initially the dominant opinion $w=1$. Notice however that in the two-against-one model we observe in any case the emergence of a consensus on the initially dominant opinion, cf. Figure~\ref{fig:mean}a;
\item For $\alpha_q=2.25$ we obtain $m_{C,q}=240$, $m_C=124.5$ and $\sigma_C\approx 276$, whence $\rho_{WC}^0\approx -0.28$. This time the first condition in~\eqref{eq:2-vs-1.phase_trans} is fulfilled, indeed $\rho_{WC}^0\approx-0.28<-\frac{m_C}{\sigma_C}\cdot\frac{m_W^0}{\sqrt{1-{(m_W^0)}^2}}\approx -0.2$. Therefore, we observe polarisation switch in both the two-against-one model (together with emergence of consensus), cf. Figure~\ref{fig:mean}a, and its Ochrombel-type simplification (without emergence of consensus), cf. Figure~\ref{fig:mean}b. A suitable reduction of the Pareto index $\alpha_q$ has made the tail of the connectivity distribution $K_{\alpha_q,\beta_q}$ heavy enough to produce a sufficient number of influencers expressing initially the opinion opposite to the dominant one. This is also confirmed by the mean connectivity $m_{C,q}$, which in this case is consistently larger than $m_{C,p}$.
\end{enumerate}

Figure~\ref{fig:mean} shows the time trend of the mean opinion $m_W$ of the two-against-one model (panel a) and of the Ochrombel-type simplification (panel b) in the two cases discussed above. Solid lines are the graphs of the functions $t\mapsto m_W(t)$ obtained from the kinetic description. Those in panel a are obtained from the numerical solution of~\eqref{eq:2-vs-1.mW} by means of a fourth-order Runge-Kutta method while those in panel b are plotted out of the analytical expression~\eqref{eq:Ochrombel.mW}. Markers indicate instead the means of the Monte Carlo solutions of the corresponding stochastic particle models~\eqref{eq:2-vs-1.particle}-\eqref{eq:2-vs-1.Theta} and~\eqref{eq:Ochrombel.particle}-\eqref{eq:Ochrombel.Theta} with initial conditions sampled from the distribution~\eqref{eq:f0}.

\begin{figure}[!t]
\centering
\includegraphics[width=0.5\textwidth]{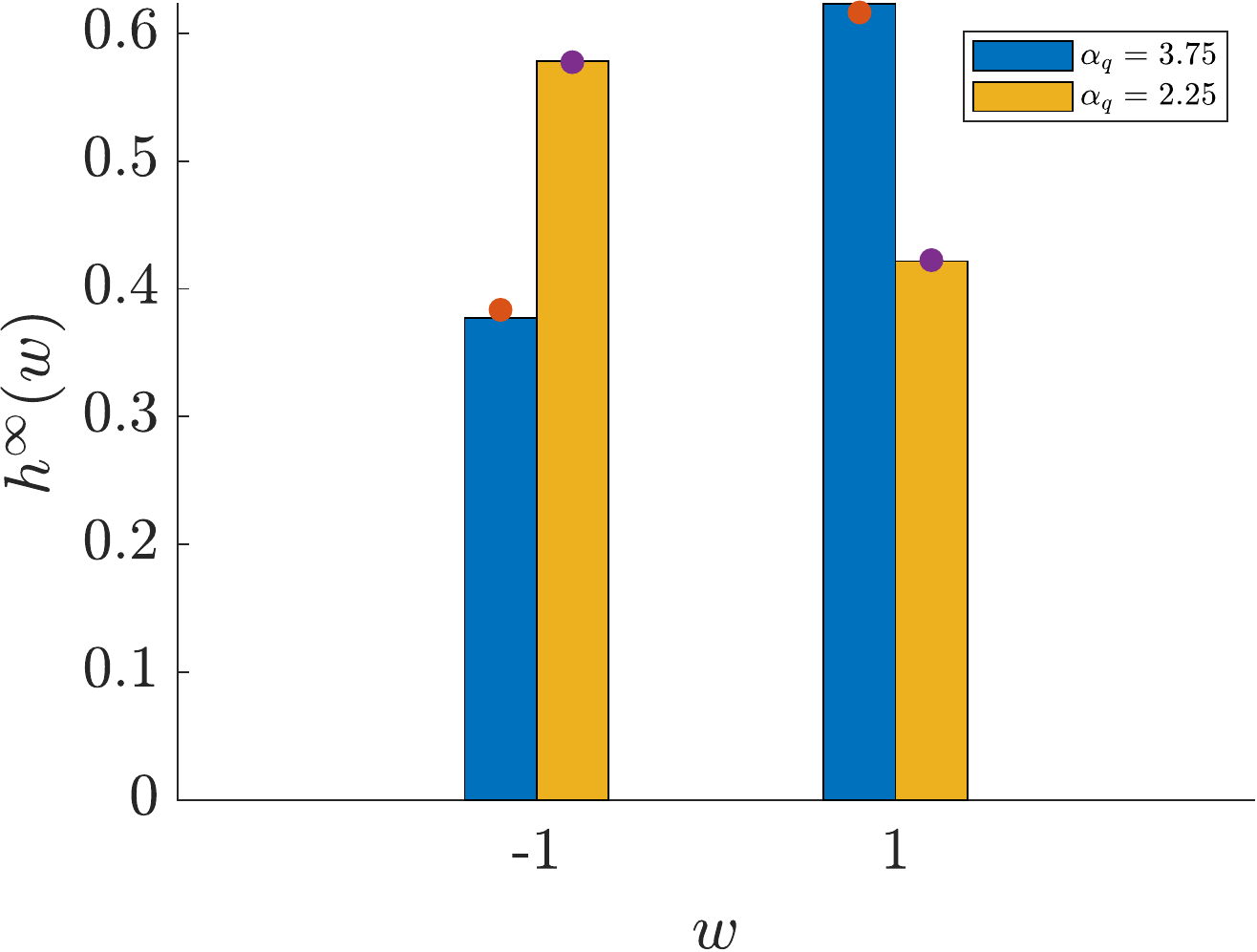}
\caption{Asymptotic opinion distribution of the Ochrombel simplification for the values of the parameters discussed in the text. Histogram bars are the Monte Carlo solutions of the stochastic particle models whereas markers are the theoretical values obtained from the kinetic description.}
\label{fig:hist}
\end{figure}

Figure~\ref{fig:hist} shows instead the asymptotic opinion distribution $h^\infty$ emerging in the Ochrombel-type simplification. Histogram bars are computed out of the Monte Carlo solution of the stochastic particle model. Markers indicate instead the values predicted by the theory, cf.~\eqref{eq:Ochrombel.hinf}.

\section{Conclusions}
\label{sect:conclusions}
In this paper, we have proposed a strategy to study opinion formation on social networks, in particular the emergence of polarisation switch, which takes advantage of a statistical description of the network embedded into a kinetic description of the opinion dynamics of social network users. Unlike other approaches, this has allowed us to address very general connectivity distributions not confined to the cases of complete graphs or regular lattices. Our main idea consists in assuming that the connectivity of a user determines the probability that their opinion reaches and influences the opinion of another user. This gives rise to non-Maxwellian kinetic equations for the joint opinion-connectivity distribution, in which the non-constant collision kernel depends on the connectivity. We have focused our analysis on simple opinion exchange rules in a discrete setting inspired by the celebrated Sznajd model~\cite{sznajd-weron2000IJMP} and its simplification proposed by Ochrombel~\cite{ochrombel2001IJMP}. Moreover, we have assumed a specific dependence of the interaction probability on the connectivity of the individuals. Interesting developments may address more general opinion exchange models, possibly in a continuous setting taking into account both consensus and dissent among the individuals~\cite{loy2020CMS,toscani2006CMS}, and a sufficiently generic dependence of the interaction probability, hence of the collision kernel, on the connectivity of the social network users.

We stress that a polarisation switch is different from a phase transition, also frequently studied in opinion dynamics, in that it allows for a change of sign of the mean opinion usually not observed in opinion models inspired by the Sznajd one. In our model, $m_W=0$ is in general \textit{not} an equilibrium value of the mean opinion unless $m_{WC}\equiv 0$, which explains why in general the state $m_W=0$ can be crossed in time towards an asymptotic sign of $m_W$ opposite to the initial one. In particular, from the evolution equations of $m_W$ reported in Section~\ref{sect:polarisation_switch} and from the further considerations proposed in Section~\ref{sect:stat.indep}, it is clear that the presence of a social network featuring a non-zero correlation between the opinion and the connectivity of the users plays a crucial role in the possible appearance of a polarisation switch.

\section*{Acknowledgements}
This research was partially supported  by the Italian Ministry for Education, University and Research (MIUR) through the ``Dipartimenti di Eccellenza'' Programme (2018-2022), Department of Mathematical Sciences ``G. L. Lagrange'', Politecnico di Torino (CUP: E11G18000350001).

NL's postdoctoral fellowship is funded by INdAM (Istituto Nazionale di Alta Matematica ``F. Severi'', Italy).

NL and AT are members of GNFM (Gruppo Nazionale per la Fisica Matematica) of INdAM, Italy.

\bibliographystyle{plain}
\bibliography{LnRmTa-polarisation_switch}
\end{document}